# One-dimensional radiation-hydrodynamic scaling studies of imploding spherical plasma liners

T. J. Awe*, C. S. Adams, J. S. Davis, D. S. Hanna, S. C. Hsu

*Physics Division, Los Alamos National Laboratory, Los Alamos, NM 87545*

J. T. Cassibry

*University of Alabama, Huntsville, Huntsville, AL 35899*

*awetj@lanl.gov

One-dimensional radiation-hydrodynamic simulations are performed to develop insight into the scaling of stagnation pressure with initial conditions of an imploding spherical plasma shell or "liner." Simulations reveal the evolution of high-Mach-number ($M$), annular, spherical plasma flows during convergence, stagnation, shock formation, and disassembly, and indicate that cm- and $\mu$s-scale plasmas with peak pressures near 1 Mbar can be generated by liners with initial kinetic energy of several hundred kilo-joules. It is shown that radiation transport and thermal conduction must be included to avoid non-physical plasma temperatures at the origin which artificially limit liner convergence and thus the peak stagnation pressure. Scalings of the stagnated plasma lifetime ($\tau_{stag}$) and average stagnation pressure ($P_{stag}$, the pressure at the origin, averaged over $\tau_{stag}$) are determined by evaluating a wide range of liner initial conditions. For high-$M$ flows, $\tau_{stag} \sim \Delta R / v_0$, where $\Delta R$ and $v_0$ are the initial liner thickness and velocity, respectively. Furthermore, for argon liners, $P_{stag}$ scales approximately as $v_0^{15/4}$ over a wide range of initial densities ($n_0$), and as $n_0^{1/2}$ over a wide range of $v_0$. The approximate scaling $P_{stag} \sim M^{3/2}$ is also found for a wide range of liner-plasma initial conditions.

## I: INTRODUCTION

Imploding spherical plasma shells, or "liners," formed by an array of convergent high-Mach-number ($M$) plasma jets, are potentially attractive for forming cm- and μs-scale high-energy-density (HED) plasmas for scientific studies and as a standoff driver magneto-inertial fusion [1,2,3,4] (MIF). The Plasma Liner Experiment (PLX) [5] at Los Alamos National Laboratory (LANL) plans to merge thirty high-$M$ dense plasma jets in spherically-convergent geometry to explore the feasibility of forming imploding spherical plasma liners. Pulsed plasma guns of modest size (~1 m long) can generate plasma jets with $n$~$10^{17}$ cm$^{-3}$, $v$~50 km/s, and $M$~10-35 [6], where $n$ and $v$ are the plasma number density and velocity, respectively, at the gun muzzle. Discrete jets coalesce at the merging radius ($R_m$) to form an imploding spherical plasma liner. The liner propagates inward and stagnates at the origin ("void closure"), where a strong shock is launched radially outward, creating high-pressure post-shocked plasma. Physics issues associated with single-jet propagation (e.g., jet expansion, cooling, atomic physics effects, and stability), multi-jet merging (e.g., oblique shock formation and heating with associated $M$ downshift), and the subsequently formed plasma liner (e.g., liner uniformity and stability, pressure amplification during spherical convergence, shock formation and stagnation, and size and lifetime of the stagnated plasma) will be examined by PLX. Examples of fundamental HED plasma physics research that will be enabled by PLX include studies of magnetized HED plasma transport and stability, astrophysically-relevant jet formation via rotating plasma disks [7], collisionless shocks via the head-on collision of two plasma jets, and atomic physics studies of high-charge-state ions in density/temperature regimes lacking validated equation-of-state (EOS) and opacity models. In addition, successful demonstration of imploding plasma liner formation will enable further explorations of this concept as a versatile standoff MIF driver.

The primary purpose of this work is to gain insight into the scaling of the post-shock-plasma stagnation pressure and lifetime with liner initial conditions using computationally inexpensive one-dimensional (1D) radiation-magnetohydrodynamic (R-MHD) simulations. This work assumes that a spherically symmetric imploding plasma liner has already been formed at $R_m$ and does not treat the jet merging process, target plasma compression, nor multi-dimensional effects. (Three-dimensional (3D) jet merging and liner implosion physics are also being studied using 3D ideal hydrodynamics [8] and particle-in-cell [9] codes.). Simulations with PLX-relevant liner initial conditions will guide experimental campaigns, while scaling studies of higher-energy liners are critical for further evaluation of the plasma-liner concept for HED physics and MIF applications. 1D R-MHD simulations (radiation pressures remain well below material pressures, but radiation is an important energy transport mechanism; for reasons explained in Sec. II, magnetic fields are not included) using the RAVEN [10] and HELIOS [11] codes provide insight into the evolution of converging plasma liners. Simulations indicate that liner stagnation pressures of ~1 Mbar can be sustained for ~1 µs using PLX-relevant liner initial conditions (kinetic energy ~300 kJ). The simulations presented in this paper provide a likely upper bound (since multi-dimensional effects are ignored) on stagnation pressure scaling with initial liner density ($n_0$) and velocity ($v_0$). In addition, the effect of variations in liner dimensions, total energy, adiabatic index, atomic species, and initial temperature, as well as effects of different radiation transport, thermal conduction, and EOS treatments, are reported.

The remainder of this article is organized as follows. Section II describes the modeling approach and specifies how simulation parameters have been selected. In Sec. III, brief descriptions of the RAVEN and HELIOS codes are given, and verification test results are provided. Section IV qualitatively describes the evolution of the high-$M$ liners considered in this study. Section V shows the importance of including radiation transport and thermal conduction in simulations. Section VI includes dwell time and stagnation pressure scaling results, a discussion of adiabatic index and atomic mass effects, and simulation results obtained using a non-local-thermodynamic-equilibrium (non-LTE) Ar EOS table. Finally, Sec. VII gives concluding remarks.

**II: MODELING APPROACH**

All simulations begin by assuming a spherical plasma (typically argon) liner (Fig. 1) of uniform initial density ($\rho_0$ or number density $n_0$) and temperature ($T_0$) propagates toward the origin with uniform initial speed ($v_0$). The liner's initial inner radius is $r_{in}=R_m$ where the merging radius, $R_m$, is the radius at which the leading edge of the discrete plasma jets meet. The initial outer radius is $r_{out}=R_m+\Delta R$ where $\Delta R$ is the liner thickness. The liner edges are assumed initially sharp, with the transition from plasma to vacuum across a single cell. While not used as input parameters in 1D simulations, the chamber radius, ($R_c=137.2$ cm for PLX), the total number of jets ($N=30$), and the initial jet cross-sectional radius ($r_0=5.0$ cm) are used when calculating $R_m$ and $\Delta R$.

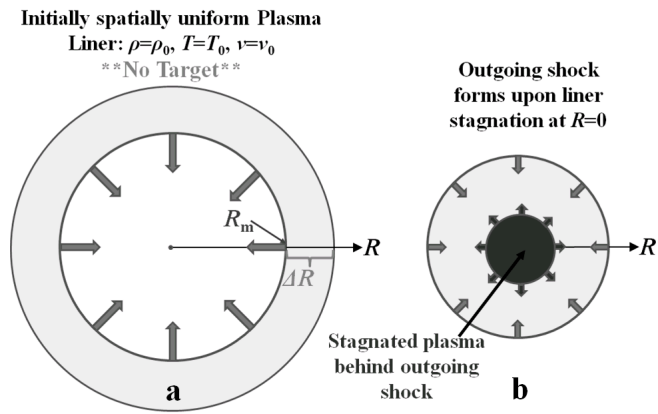

FIG. 1: (a) Initial configuration used in 1D plasma liner simulations. Plasma with constant density, temperature, and fluid velocity extends from $r_{in}=R_m$ to $r_{out}=R_m+\Delta R$. Simulations contain no target plasma. (b) The plasma configuration after collapse upon the origin. A spherical shock wave propagates outward into the remaining incoming liner plasma. Behind the shock front, high-pressure stagnated plasma persists until the shock front and outer edge of the liner meet.

For dense, cool plasma jets ($n\sim10^{16}$–$10^{17}$ cm$^{-3}$, $T<3$ eV) with *initially* modest levels of embedded magnetic field ($B<1$ kG), the collision frequency is much greater than the gyro-frequency for both electrons and ions, and therefore magnetic field and MHD effects have been ignored in this study. Furthermore, the time for the jet to propagate to the merging radius (~20 μs) exceeds the resistive diffusion time for the field (~17 μs assuming $T$=2 eV, $Z_{eff}$=1, and 5 cm length scale), and thus any initial field will decay to even more inconsequential values by the time the jets reach $R_m$.

Two simulation series with separate general constraints are explored. The first series examines the liner parameter space accessible to PLX. The second series explores higher-energy liners, including those envisioned to achieve MIF-relevant pressures. Each series is discussed separately in the following subsections.

## A. Series I—The PLX parameter space

A series of simulations has been designed to explore PLX-accessible liner parameters. Changes in stagnated plasma pressure ($P_{stag}$) and lifetime ($\tau_{stag}$) due to variations in liner initial density ($\rho_0$), velocity ($v_0$), geometry, and kinetic energy ($KE_0$) are explored. The following assumptions are made:

1.  The initial plasma temperature ($T_0$) is 2.8 eV. PLX initial liner temperatures are not yet known precisely; jets may cool to 1 eV or less prior to liner formation, but the plasma may then be reheated by shock formation upon jet merging. RAVEN simulations use a 3 temperature (3T) model, and set the initial ion, electron, and radiation temperatures to $T_{i,0}=T_{e,0}=T_{r,0}=T_0$=2.8 eV. HELIOS includes 1T and 2T ($T_i \neq T_e$) temperature models; however, only the 1T model is available when using an ideal-gas EOS. HELIOS results presented in Sec. VI.E use a PROPACEOS [12] EOS table [13] and the 2T option.

2.  Jets propagate from the chamber radius ($R_c$=137.2 cm) to $R_m$ in a time $t_{prop}$=$(R_c - R_m)/v_0$, where $v_0$ is the initial jet velocity ($t_{prop}$ is used to estimate radial jet expansion in $R_m$ calculations)

3.  The jet radius expands at constant sound speed ($c_s$), *i.e.,* $r_j(t)$=$r_{j,0}+c_s t$. The validity of this assumption improves with increasing $M$. Adiabatic expansion models, which include plasma cooling, result in reduced jet expansion and increased $M$ and $R_m$ for given $R_c, r_{j,0}$, and $T_0$.

The merging radius $R_m$ is calculated by setting the total surface area of all (cylindrically symmetric) expanded jets to the surface area of a sphere with radius $r = R_m$. It can be shown that $R_m$=$(r_0+Q \cdot R_c)/(Q+2 \cdot N^{-1/2})$ where $Q \equiv c_s/v$=$1/M$. Initial velocities of 50 and 100 km/s (achievable with small-scale plasma guns) are chosen, resulting in merging radii of $R_m$=0.329 m and $R_m$=0.241 m, respectively. For 2.8 eV and 1.0 eV Ar plasmas, kinetic energy exceeds internal energy for velocities above 4.5 km/s and 2.7 km/s, respectively; thus the initial liner energy is mostly kinetic.

Finally, by choosing $KE_0$, for a given $v_0$ and $\rho_0$, the plasma mass is defined. Experimentally achievable number densities (at $R_m$) of $10^{15}$ cm$^{-3}$ and $10^{16}$ cm$^{-3}$ are chosen. Then, with $R_m$ known, the liner thickness ($\Delta R$) is readily calculated. The total liner $KE_0$ will be a fraction of the total capacitively stored energy of the PLX pulsed power system. Each plasma gun will use six 60 kV, 6.0 µF capacitors in parallel, with a maximum total stored energy of ~65 kJ. A system of 30 guns can then store ~2.0 MJ. Gun electrical efficiency of 25% is realistic, but since experiments will typically be conducted with capacitors at less than full charge, $KE_0$ is set to either 150 or 300 kJ. Table 1 summarizes the initial conditions for 8 simulations, all with PLX-accessible liner parameters.

---

**Table 1:** Initial liner parameters for the series of "PLX-accessible" simulations. Cases 1, 3, 5, and 7 differ only by a factor of two in $KE_0$ compared to cases 2, 4, 6 and 8, respectively (accomplished by changing $\Delta R$ and therefore also the total liner mass). All simulations assume $T_0$=2.8 eV argon plasma, and an ideal gas EOS with $\gamma$=5/3 ($\gamma$ is the adiabatic index). Select results (discussed in detail in Sec. VI), including the stagnation pressure and time ($P_{stag}$ and $\tau_{stag}$, defined precisely in Sec. IV) and maximum pressure ($P_{max}$) are included in Table 1 (columns 9–13) for reference. For further reference, liner performance parameters (discussed in Section VI.C) are included in the final two columns.

| Case | $v_0$ (km/s) | $n_0$ (cm$^{-3}$) | $\rho_0$ (kg/m$^3$) | $KE_0$ (J) | $R_m$ [m] | Mass [mg] | $\Delta R$ [m] | $\tau_{stag}$ (µs) | $P_{stag}$ [Pa] | $P_{max}$ [Pa] | $P_{stag} \times \tau_{stag}$ | $(P_{stag} \times \tau_{stag})/KE_0$ |
|------|------|------|------|------|------|------|------|------|------|------|------|------|
| PLX 1 | 50 | 1.0E+15 | 6.63E-5 | 1.5E+5 | 0.329 | 120 | 0.447 | 8.18 | 1.87E+9 | 6.66E+10 | 1.53E+4 | 0.10 |
| PLX 2 | 50 | 1.0E+15 | 6.63E-5 | 3.0E+5 | 0.329 | 240 | 0.636 | 12.09 | 1.69E+9 | 5.13E+10 | 2.05E+4 | 0.07 |
| PLX 3 | 50 | 1.0E+16 | 6.63E-4 | 1.5E+5 | 0.329 | 120 | 0.099 | 1.60 | 2.29E+10 | 4.83E+11 | 3.66E+4 | 0.24 |
| PLX 4 | 50 | 1.0E+16 | 6.63E-4 | 3.0E+5 | 0.329 | 240 | 0.166 | 3.08 | 1.60E+10 | 4.63E+11 | 4.93E+4 | 0.16 |
| PLX 5 | 100 | 1.0E+15 | 6.63E-5 | 1.5E+5 | 0.241 | 30 | 0.255 | 2.42 | 3.45E+10 | 6.04E+12 | 8.35E+4 | 0.56 |
| PLX 6 | 100 | 1.0E+15 | 6.63E-5 | 3.0E+5 | 0.241 | 60 | 0.372 | 3.59 | 2.37E+10 | 4.1E+12 | 8.51E+4 | 0.28 |

| PLX 7 | 100 | 1.0E+16 | 6.63E-4 | 1.5E+5 | 0.241 | 30 | 0.051 | 0.40 | 8.12E+11 | 1.82E+13 | 3.25E+5 | 2.16 |
| PLX 8 | 100 | 1.0E+16 | 6.63E-4 | 3.0E+5 | 0.241 | 60 | 0.089 | 0.81 | 4.64E+11 | 1.86E+13 | 3.76E+5 | 1.25 |

## B. Series II—Plasma liner scaling studies

An additional series of 16 simulations (parameters are given in Table 2) is used to evaluate how stagnated plasma parameters scale over a wide range of initial liner conditions (e.g., with $\rho_0$, $v_0$), from PLX-accessible liners to those with much higher energies. Some repetition in $KE_0$ exists, allowing a more direct comparison between simulations. All simulations consider liners with $R_m$=0.241 m and $\Delta R$=0.255 m (that of PLX 5 in Table 1), and most use Ar plasma, an ideal gas EOS with $\gamma$=5/3, and $T_0$=1.0 eV. Additional simulations use the same parameters as given in Table 2 but with different $\gamma$, atomic species (and thus number density), and/or $T_0$. Initial conditions used in these additional simulations will be clearly defined when their results are presented.

**Table 2:** Initial liner parameters for stagnation pressure ($P_{stag}$) scaling studies. All simulations use $R_m$=0.241 m and $\Delta R$=0.255 m. Furthermore, simulations assume $T_0$=1.0 eV argon plasma and use an ideal gas EOS with $\gamma$=5/3, unless otherwise specified. Select results (columns 6–10) are also presented (see Table 2 caption for details).

| Run | $n_0$ (cm$^{-3}$) | $\rho_0$ (kg/m$^3$) | $v_0$ [km/s] | $KE_0$ [J] | $\tau_{stag}$ [μs] | $P_{stag}$ [Pa] | $P_{max}$ [Pa] | $P_{stag} \times \tau_{stag}$ | $(P_{stag} \times \tau_{stag})/KE_0$ |
|---|---|---|---|---|---|---|---|---|---|
| 1 | 2.5E+15 | 1.66E-4 | 25 | 2.35E+4 | 8.82 | 2.37E+8 | 3.17E+9 | 2.09E+3 | 0.09 |
| 2 | 2.5E+15 | 1.66E-4 | 50 | 9.39E+4 | 4.51 | 3.11E+9 | 8.58E+10 | 1.40E+4 | 0.15 |
| 3 | 2.5E+15 | 1.66E-4 | 100 | 3.76E+5 | 2.29 | 3.43E+10 | 8.82E+11 | 7.84E+4 | 0.21 |
| 4 | 2.5E+15 | 1.66E-4 | 200 | 1.50E+6 | 1.03 | 6.29E+11 | 4.26E+13 | 6.48E+5 | 0.43 |
| 5 | 1.0E+16 | 6.63E-4 | 25 | 9.39E+4 | 8.84 | 4.78E+8 | 4.41E+9 | 4.23E+3 | 0.05 |

| | | | | | | | | |
|---|---|---|---|---|---|---|---|---|
| 6 | 1.0E+16 | 6.63E-4 | 50 | 3.76E+5 | 4.56 | 6.79E+9 | 1.32E+11 | 3.09E+4 | 0.08 |
| 7 | 1.0E+16 | 6.63E-4 | 100 | 1.50E+6 | 2.33 | 8.39E+10 | 2.40E+12 | 1.95E+5 | 0.13 |
| 8 | 1.0E+16 | 6.63E-4 | 200 | 6.01E+6 | 1.19 | 8.32E+11 | 2.35E+13 | 9.86E+5 | 0.16 |
| 9 | 4.0E+16 | 2.65E-3 | 25 | 3.76E+5 | 9.11 | 9.72E+8 | 6.49E+9 | 8.85E+3 | 0.02 |
| 10 | 4.0E+16 | 2.65E-3 | 50 | 1.50E+6 | 4.61 | 1.47E+10 | 1.93E+11 | 6.77E+4 | 0.05 |
| 11 | 4.0E+16 | 2.65E-3 | 100 | 6.01E+6 | 2.29 | 1.79E+11 | 3.43E+12 | 4.10E+5 | 0.07 |
| 12 | 4.0E+16 | 2.65E-3 | 200 | 2.40E+7 | 1.17 | 2.38E+12 | 1.04E+14 | 2.78E+6 | 0.12 |
| 13 | 1.6E+17 | 1.06E-2 | 25 | 1.50E+6 | 9.33 | 2.01E+9 | 1.00E+10 | 1.88E+4 | 0.01 |
| 14 | 1.6E+17 | 1.06E-2 | 50 | 6.01E+6 | 4.76 | 3.03E+10 | 2.44E+11 | 1.44E+5 | 0.02 |
| 15 | 1.6E+17 | 1.06E-2 | 100 | 2.40E+7 | 2.37 | 3.38E+11 | 4.25E+12 | 7.99E+5 | 0.03 |
| 16 | 1.6E+17 | 1.06E-2 | 200 | 9.62E+7 | 1.19 | 5.21E+12 | 1.36E+14 | 6.17E+6 | 0.06 |

## III: CODE DESCRIPTIONS AND VERIFICATION TESTS

Two codes have been used to simulate plasma liner implosions in this work. Cross-code comparisons enable direct evaluation of numerical algorithms and EOS/transport models, and give increased confidence in each code's results. The majority of the results in this manuscript are from RAVEN simulations, but results from HELIOS, including simulations using the PROPACEOS non-local-thermodynamic-equilibrium (non-LTE) Ar EOS table, are also reported.

RAVEN [10] is a 1D (plane, cylindrical, or spherical symmetry) radiation-magnetohydrodynamic (R-MHD) Lagrangian code with multiple EOS options, including ideal gas and SESAME table lookup; however, readily available Ar SESAME tables (which go down to a density of $10^{17}$ cm$^{-3}$) do not cover the full range of densities needed to model plasma liner implosions, and were therefore not used in this study. Results using recently generated PROPACEOS non-LTE EOS (valid down to a density of $10^{10}$ cm$^{-3}$), which do cover the needed density range, will be reported in a separate forthcoming paper. In RAVEN, multiple materials can exist within a single module; the materials can be in direct contact or coupled across an internal- and kinetic-energy-free vacuum region. RAVEN includes both 1 temperature (1T) and 3 temperature (3T) models. Here, the 3T model is used exclusively, allowing separate electron, ion, and radiation temperatures ($T_i$, $T_e$, and $T_r$). Use of the 3T model allows a more accurate representation of the varying importance of radiation; the radiation field is initially decoupled from low-density plasma, but then as the density increases, the plasma becomes optically grey or thick, $T_e$ and $T_r$ equilibrate, and radiation can play an important role in plasma energetics (see Sect. V). Furthermore, allowing $T_i \neq T_e$ enables the shock to preferentially heat the more massive ions. Therefore, immediately behind the shock front $T_i \neq T_e$; however, the electron-ion equilibration time is typically quite short in comparison to other time scales of interest in the problem. In RAVEN's energy balance equations, $T_i$ and $T_r$ are coupled only through their interactions with electrons. Separate thermal conductivities are used for electrons and ions, and radiation-transport calculations use either Rosseland or Planckian opacities.

HELIOS-CR [11] is a 1D (plane, cylindrical, or spherical symmetry) R-MHD Lagrangian code with multiple EOS options, including ideal gas, SESAME table lookup, PROPACEOS EOS/multi-group opacity data tables, and non-LTE properties computed using in-line collisional-radiative modeling. HELIOS-CR allows multiple regions composed of different materials in different states. While absolute vacuum regions cannot be specified, "voids" which exclude joule heating and magnetic field contributions to the pressure terms in the momentum equation may be included. The momentum equation is solved for a single fluid, with pressure contributions from electrons, ions, radiation, and magnetic field. Energy transport is treated with either a one-temperature ($T_i = T_e$) or two-temperature ($T_i \neq T_e$) thermal conduction model based on Spitzer conductivities (unless otherwise specified). Radiation transport calculations reference the electron temperature, with emission and absorption determined from opacity table data using multiple frequency groups.

Each code's ability to simulate strong shocks in a spherically convergent geometry, and in particular, proper implementation of an artificial viscosity for shock front simulation, has been tested against the analytic solutions to the "Noh problem" [14], which considers an infinite, isotropic, zero temperature (infinite $M$) fluid with an initially-uniform radially-inward flow velocity. A shock is generated at the origin and propagates outward. Analytic solutions exist for the post-shock pressure ($P_s$), density ($\rho_s$), and temperature ($T_s$), along with the shock-front velocity ($v_s$), as functions of the initial velocity ($v_0$), density ($\rho_0$), and $\gamma$. For a spherically symmetric $\gamma=5/3$ hydrogen plasma with $\rho_0=0.0166$ kg/m$^3$ ($n_0=10^{19}$ cm$^{-3}$) and $v_0=100$ km/s (radially inward), the following solutions hold:

$$\rho_s = \rho_0 \left( \frac{\gamma+1}{\gamma-1} \right)^3 = \rho_0 \left( \frac{8/3}{2/3} \right)^3 = 64 \rho_0 \sim 1.06 \; kg/m^3 \tag{1}$$

$$P_s = \rho_0 v_0^2 \frac{(\gamma+1)^3}{2(\gamma-1)^2} = \frac{64}{3} \rho_0 v_0^2 = 3.54 \times 10^9 \; Pa \tag{2}$$

$$v_s = -\frac{1}{2} v_0 (\gamma-1) = 33.33 \; km/s \tag{3}$$

Then, by the ideal gas law:

$$T_s = \frac{m}{k_B} \frac{P_s}{\rho_s} = (1.203 \times 10^{-4}) \frac{P_s}{\rho_s} \sim 34.5 \; eV \tag{4}$$

Each code has been verified against the Noh problem. Normalized pressure, density, and temperature results are plotted versus the number of numerical grid cells used (Fig. 2). Results converge toward analytic solutions, with the exception of the RAVEN-calculated temperature, which, while convergent, approaches $T \sim 1.12 \cdot T_s$ (the deviation from the analytic solution is likely due to "excess wall heating on shock formation" as described in [14]. The level of error is considered acceptable for the purpose of this study). The shock speed ($v_s$) is observed to be nearly constant and in agreement with the analytic solution. For RAVEN simulations with 1200, 2400, and 4800 cells, during the first 200 ns, the average $v_s$ is found to equal 35, 34, and 35 km/s, respectively.

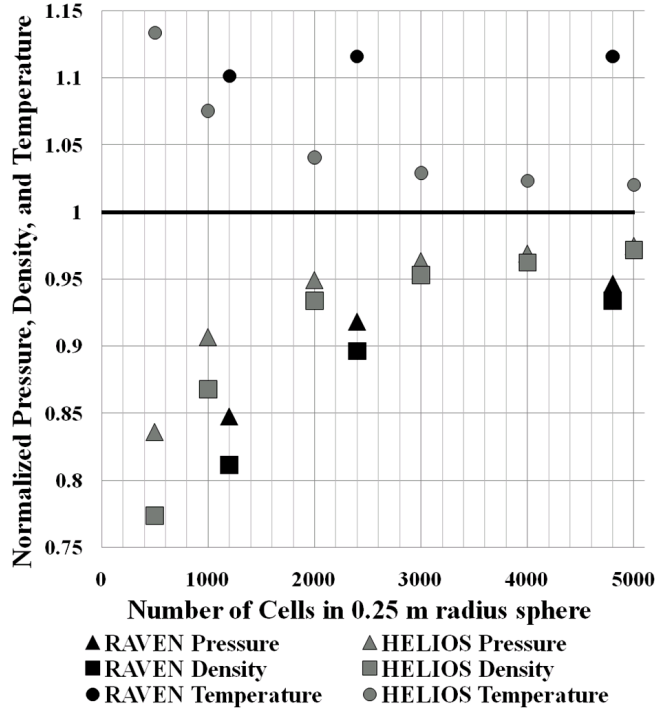

**RAVEN Pressure** ▲ **HELIOS Pressure**
■ **RAVEN Density** ■ **HELIOS Density**
● **RAVEN Temperature** ● **HELIOS Temperature**

FIG. 2: RAVEN and HELIOS normalized pressure, density, and temperature versus number of cells in a 0.25 m radius sphere for the Noh problem verification tests. Convergence is shown for all calculations. All solutions converge toward a value nearly equal to the analytic solution, with the exception of the RAVEN-calculated temperature, which approaches $T \sim 1.12 \cdot T_s$.

Grid resolution convergence tests using the annular plasma liner geometry shown in Fig. 1 find that cell size has the most profound effect on the dynamics of the leading edge of the liner during convergence. In particular, higher grid resolution results in increased expansion at the liner's leading edge. Liner material therefore reaches the origin somewhat earlier, but the lasting effect is minimal; time averaged stagnated plasma pressures near the origin increase somewhat for higher-resolution simulations, but stagnated plasma lifetimes are nearly identical. While a precise examination of leading-edge physics requires high resolution, for the present study, which is most concerned with the scaling of averaged or integrated parameters such stagnation pressure and lifetime, high resolution is not critical. The majority of the results reported considered liners with an initially uniform (Lagrangian) grid of either 250 or 500-μm-thick cells.

## IV: PLASMA LINER EVOLUTION

The qualitative evolution of high-$M$ imploding plasma liners is similar over a wide range of initial conditions. Fig. 3 displays the temporal evolution of the cell that is initially 1.0 cm from the leading edge of the liner ($r_0=R_m+1.0$ cm) for case 6 of Table 2, while Fig. 4 plots separately the pressure, density, electron temperature, and fluid velocity versus radius across the liner at $t$=5.0, 6.5, 8.0, 9.5, 10.0, and 10.5 μs. The imploding plasma liner behaves approximately as a steady-state convergent flow until reaching the origin, where an outwardly propagating shock forms. High-density stagnated plasma persists behind the shock until the shock front meets the trailing edge of the liner, at which point disassembly occurs.

Prior to the leading edge of the liner reaching the origin (from $t$=0 to $t$~5 μs), the plasma exhibits quasi-steady-state behavior, with spherical convergence resulting in increased pressure and density. In the liner interior, to high accuracy, the density increases as $\rho(r)=\rho_0(r_0/r)^2$ ($r_0$ and $r$ are the initial and time varying radius, respectively, of the fluid element), which has been reported earlier for high-$M$ spherically convergent flows [15,16]. Near the liner edges (vacuum interfaces), rarefaction waves initially result in reduced local density and a slightly broadened density profile. As the plasma nears the origin, the pressure, density, and temperature increase rapidly, and the average ionization state grows (the time-dependent average ionization, $Z_{eff}$, is determined by the Saha equation using calculated plasma parameters, but does not contribute to the total plasma pressure when using an ideal gas EOS). The fluid velocity begins to decrease, and steady-state solutions no longer adequately describe the fluid behavior. Plasma conditions immediately prior to shock formation are shown by the solid black curves in Fig. 4.

When the liner reaches the origin the innermost plasma achieves a peak pressure ($P_{max}$) of 1.9 Mbar at $t$=$t_{max}$=5.1 μs. A spherical shock propagates outward through the still converging trailing mass of the liner. From 6.5 μs to 9.5 μs the shock front travels approximately 4.3 mm, or at an average outward velocity $v_s$~1.4 km/s; a small fraction of the initial liner inward velocity ($v_0$=50 km/s). The volume of stagnated plasma grows and plasma pressures behind the shock front persist at nearly 0.1 Mbar for several microseconds. Then at $t$=$t_{fall}$~10 μs the rarefaction wave at the outer edge of the liner reaches the shock front (grey dotted curves, Fig. 4), and the pressure, density, and temperature of the stagnated plasma decay. The stagnated plasma is no longer confined, and begins to expand (grey dashed curves, Fig. 4). The lifetime of the high-pressure stagnated plasma is $\tau_{stag} \equiv t_{fall} - t_{max}$, and the stagnation pressure, $P_{stag}$, is defined as the plasma pressure at the origin, averaged over $\tau_{stag}$ ($P_{stag}$, and $\tau_{stag}$ pertain to stagnated liner confinement only, as no target is included in simulations). The described plasma liner evolution holds, qualitatively, for all high-$M$ liner simulations included in this report.

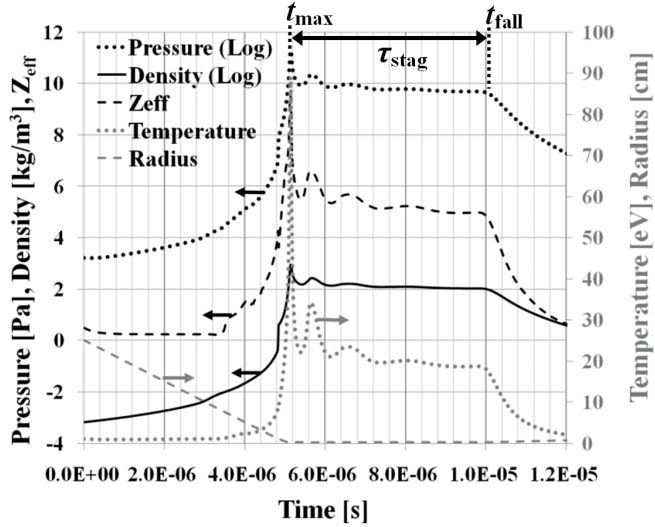

FIG. 3. RAVEN calculated pressure [Pa], density [kg/m$^3$], electron temperature [eV], average ionization state ($Z_{eff}$), and radius [cm], plotted as functions of time, for the Lagrangian cell initially at $r_0 = (R_m + 1.0)$ cm for simulation 6 of Table 2. Data is taken from a simulation with an initial cell thickness of 100 μm. Pressure and density are plotted on a logarithmic scale. Arrows and curve color indicate whether data correspond to the primary (black) or secondary (grey) vertical axis. The stagnation lifetime ($\tau_{stag}$, discussed extensively in Section VI.A) is defined as the time span between the occurrence of peak pressure ($t_{max}$) and the time of disassembly ($t_{fall}$). Disassembly occurs when the (outwardly propagating) shock front meets the (inwardly propagating) trailing edge of the liner, and is observed as the onset of the rapid decrease in pressure at the origin following the extended period of relatively constant pressure.

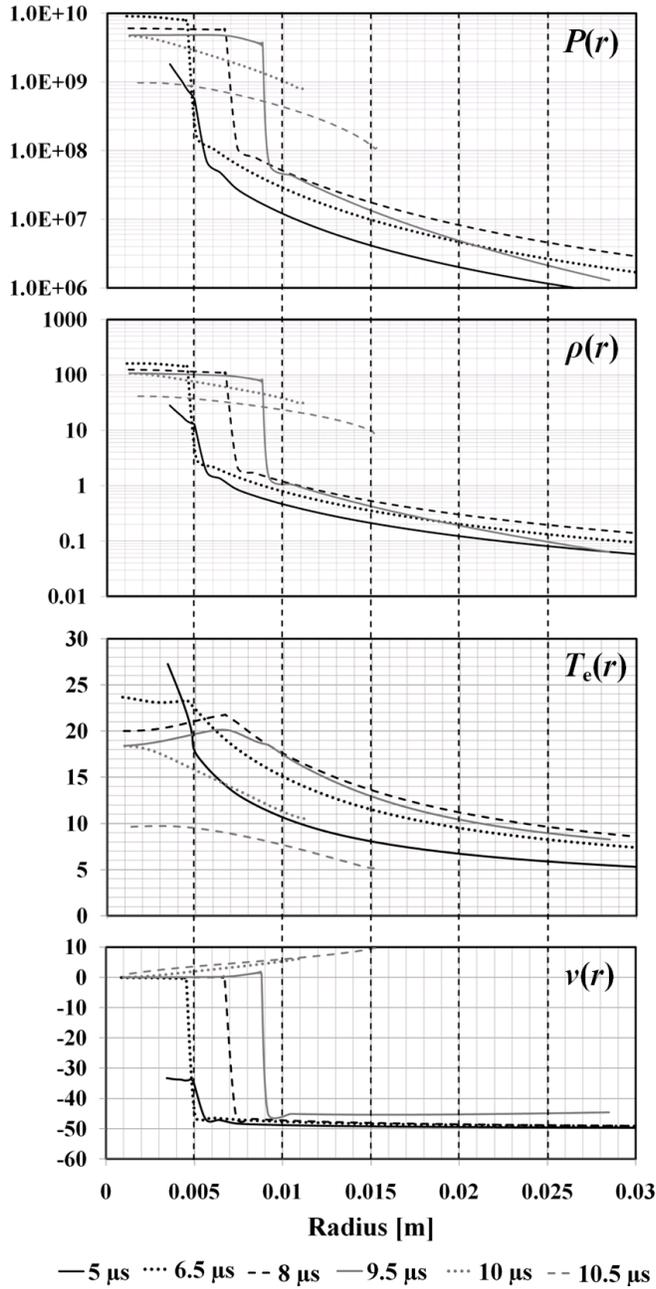

FIG. 4: RAVEN calculated pressure [Pa], density [kg/m$^3$], electron temperature [eV], and fluid velocity [km/s] as functions of radius [m] at 5.0, 6.5, 8.0, 9.5, 10.0, and 10.5 μs. Data is from the same simulation as that used to create Fig. 3. The horizontal axis is set to focus on the stagnated plasma generated near the origin, so that for t=5.0, 6.5, and 8.0 μs, the outermost portions of the liner are not shown.

## V: ROLE OF RADIATION TRANSPORT AND THERMAL CONDUCTION

Before presenting the rest of the results, it is important to note that radiation transport and thermal conduction are important physical processes in plasma-liner implosions, and do play a critical role in convergence and stagnation dynamics. Throughout the implosion process, plasma conditions vary rapidly, both in space and time. For example, at $R_m$, plasma parameters are of order $P\sim0.01$ bar, $\rho\sim10^{-4}$ kg/m$^3$, and $T\sim1$ eV, but increase to nearly 1 Mbar, $10^3$ kg/m$^3$, and up to 100's of eV upon convergence and stagnation. Furthermore, large gradients exist due to spherical convergence, rarefaction waves, and especially across the shock front. The role of radiation transport and thermal conduction vary in accord with the diverse and rapidly changing plasma parameters. Here, the role of energy transport is investigated through simulation 6 (Table 2), under the following four conditions: (1) radiation ON, thermal conduction ON; (2) radiation ON, thermal conduction OFF; (3) radiation OFF, thermal conduction ON; and (4) radiation OFF, thermal conduction OFF.

Radiation transport and thermal conduction are "turned off" in the simulations by setting the Rosseland mean free path and electron and ion thermal conductivity multipliers to zero, respectively. In RAVEN's 3T model, the ion fluid and radiation field cannot exchange energy directly but can equilibrate through separate interactions with the electron fluid, and adjustments may be made to the electron/ion equilibration time and the electron/radiation equilibration constant. While setting the Rosseland mean free path to zero does eliminate radiation transport, radiation is not eliminated; radiation temperature, pressure, and energy are still included, and the radiation field still exchanges energy with the electrons. Similarly, even when thermal conduction is suppressed, thermal energy can still be distributed by the interaction of $T_i$ with $T_e$, and of $T_e$ with $T_r$, followed by the redistribution of energy via radiation. Only when both radiation transport and thermal conduction are suppressed will these energy transport mechanisms be truly eliminated.

When energy transport is suppressed, the expanded leading edge of the liner, which is composed of a very small fraction of the total liner mass, can have a large, unphysical impact on the achieved stagnation pressure by creating a very low density yet high pressure "hot plasma bubble" of extreme temperature (e.g., 100's of keV). The trailing mass of the plasma liner stagnates upon the central bubble rather than at the origin, reducing the achieved spherical convergence and thus also the stagnation pressure (Fig. 5). Qualitatively similar results are found in HELIOS simulations. Suppressing radiation transport has a stronger effect than suppressing thermal transport, since without radiative cooling the interior expanded cells cannot collapse. In this low density, high temperature plasma, thermal conductivity is low, and heat transport therefore has a minimal effect. The role of energy transport in this plasma configuration, where optically thin (outer portions of the liner where convergence ratios are moderate), optically grey (portions of liner at high convergence ratio), and optically thick (post shock, stagnated plasma) coexist, is complex and the topic of ongoing studies. In the present work, clearly unphysical phenomena are avoided by including both radiation transport and thermal conduction in all simulations (outside of those shown in Fig. 5).

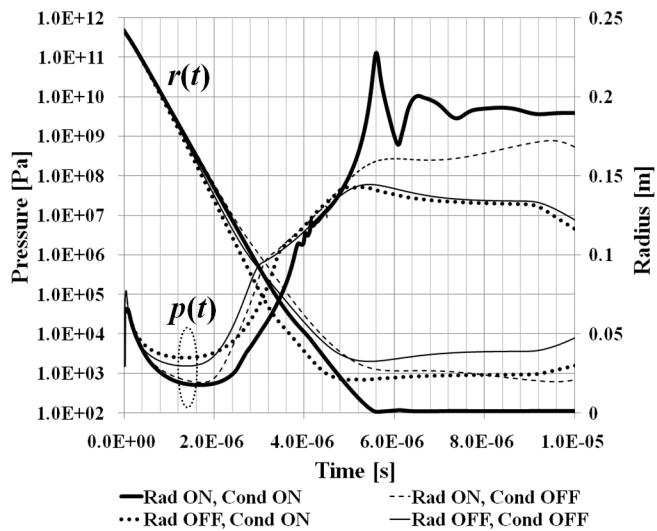

FIG. 5. $P(t)$ and $r(t)$ curves (case 6 of Table 2) for the cell initially at $r_0=(R_m+1.0)$ mm for RAVEN simulations under the following four conditions: (1) radiation ON, thermal conduction ON (bold line); (2) radiation ON, thermal conduction OFF (dashes); (3) radiation OFF, thermal conduction ON (dots); and (4) radiation OFF, thermal conduction OFF (thin line). When energy transport is suppressed, a "hot plasma bubble" forms at the origin, which reduces the achieved liner convergence ratio and stagnation pressure.

# VI: RESULTS

## A: PLX-accessible simulations—Stagnation pressure and lifetime

For high-$M$ liners, stagnation pressure increases both with $v_0$ and $\rho_0$. Fig. 6 includes $P(t)$ curves for the cell initially 1.0 mm from the leading edge of the liner ($r_0=R_m+1.0$ mm) for each of the simulations defined in Table 1. These results indicate that, for liner initial conditions realistically achievable by PLX, stagnation pressures near the origin can be sustained at over 1 Mbar for 1 µs, at 100 kBar for 2-3 µs, or at 10 kBar for over 10 µs. Correlation between liner initial conditions and stagnated plasma parameters is clear. First, peak pressure is achieved at $t_{max}\sim R_m/v_0$. Second, to be discussed in detail in Sec. VI.B, peak pressures ($P_{max}$) and sustained stagnation pressures ($P_{stag}$) increase with both $v_0$ and $\rho_0$. Third, $\Delta R$ and $v_0$ primarily determine the stagnated plasma lifetime ($\tau_{stag}$).

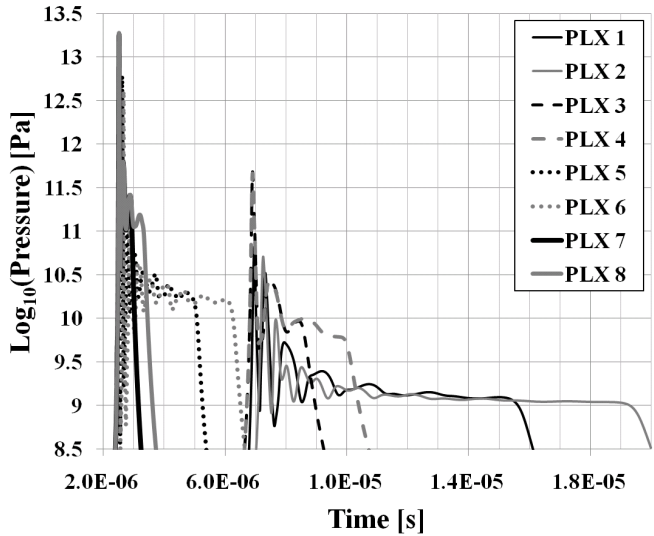

FIG. 6: RAVEN calculated $P(t)$ curves for the cell initially 1.0 mm from the leading edge of the liner ($r_0=R_m+1.0$ mm) for each of the simulations defined in Table 1. All curves are for $T_0$=1.0 eV argon, using an ideal gas EOS with $\gamma$=5/3. Cases 1, 3, 5, and 7 differ by only a factor of two in $KE_0$ compared to cases 2, 4, 6 and 8, respectively (accomplished by changing $\Delta R$ and therefore also the total liner mass; see Table 1).

While of negligible initial effect on $P(t)$, increased $\Delta R$ results in increased $\tau_{stag}$ (for fixed $v_0$ and $\rho_0$) for all high-$M$ plasma liner simulations. Disassembly occurs when the (outwardly propagating) shock front meets the (inwardly propagating) trailing edge of the liner. The stagnation time is therefore nearly equivalent to the liner thickness ($\Delta R$) divided by ($v_s - v_L$), where $v_s$ is the shock speed, and $v_L$ is the (inward) velocity of the trailing edge of the liner. In all cases examined, $v_s \ll v_L$ (see Section VI.B), and the velocity of the trailing edge of the liner remains nearly equal to the initial liner velocity ($v_L \sim v_0$) throughout the implosion. The stagnation lifetime is thus $\tau_{stag} \sim \Delta R / v_0$, as demonstrated in Fig. 7, where pressure has been plotted against the dimensionless variable $t/(\Delta R/v_0) - C_i$. (The horizontal shift, $C_i = t_{max} v_0 / \Delta R$, has been applied to each curve so that all of the pressure peaks align at zero.) On this dimensionless axis, all pressure curves fall at $t/(\Delta R/v_0) \sim 1$, showing that, to first order, $\tau_{stag} = \Delta R/v_0$.

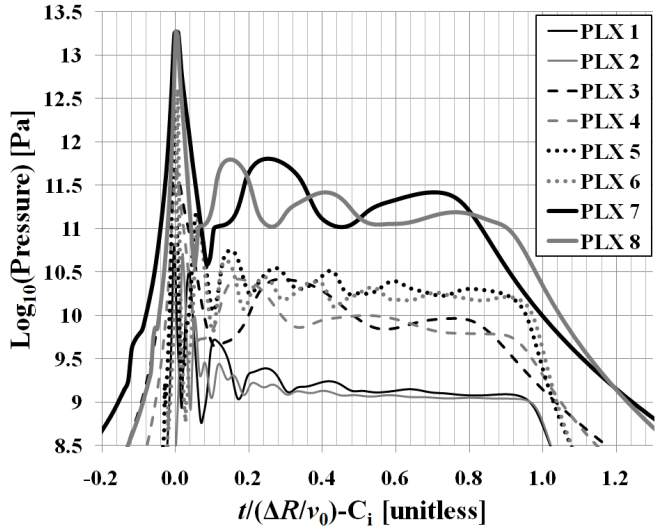

FIG. 7: Pressure data from Fig. 6 plotted versus the dimensionless variable $t/(\Delta R/v_0) - C_i$, where a horizontal shift, $C_i = t_{max} v_0 / \Delta R$, has been applied to each curve so that all of the pressure peaks align at zero. In all cases, the magnitude of $\partial p / \partial t$ of the plasma at the origin increases markedly when $t/(\Delta R/v_0) \sim 1$, implying that $\tau_{stag} \sim \Delta R/v_0$.

## B: Stagnation pressure scaling with liner density and velocity

Stagnation pressure ($P_{stag}$, defined in Sec. IV) scales with both $v_0$ and $\rho_0$, as determined by evaluating the $P(t)$ curves for each of the 16 simulations defined in Table 2 (Fig. 8). Data pertain to the cell initially located 1.0 mm from the leading edge of the liner ($r_0 = R_m + 1.0$ mm). All simulations use the same $R_m$ and $\Delta R$; therefore, the $P(t)$ curves are clearly grouped according to their initial velocity (since $v_0$ determines when the liner reaches the origin, and, for fixed $\Delta R$, also $\tau_{stag}$). For a given $v_0$, higher-density liners achieve higher $P_{stag}$. Furthermore, $P(t)$ curves of the same color share the same $KE_0$ (See Table 2). For example, all curves plotted in purple are for liners with $KE_0 = 1.5$ MJ. For a given $KE_0$, those curves with higher $v_0$ (but therefore lower $n_0$) achieve higher peak pressures (but $\tau_{stag}$ necessarily decreases).

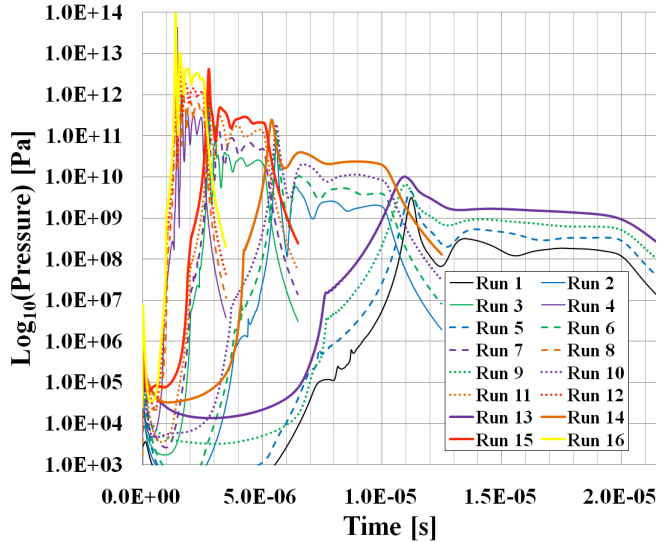

FIG. 8: (Color online) RAVEN calculated $P(t)$ curves for the cell initially located 1.0 mm from the inner edge of the liner ($r_0 = R_m + 1.0$ mm). The initial conditions for the 16 simulations included are given in Table 2. All curves are for $T_0 = 1.0$ eV argon, using an ideal gas EOS with $\gamma = 5/3$. Curves with $n_0 = 2.5 \times 10^{15}$, $1.0 \times 10^{16}$, $4.0 \times 10^{16}$, and $1.6 \times 10^{17}$ cm$^{-3}$ are plotted with thin lines, dashes, dots, and thick lines, respectively.

For a given $KE_0$, simulations show clear scaling of $P_{stag}$ with both $n_0$ and $v_0$. $P_{stag}$ is plotted versus $n_0$ for a given $v_0$ (Fig 9.a), and versus $v_0$, for a given $n_0$ (Fig 9.b). Each curve is fit to a power-law function. Over the parameter space investigated, $P_{stag}$ scales approximately as $n_0^{1/2}$ for a given $v_0$ and as $v_0^{15/4}$ for a given $n_0$. The $P_{stag}$ data used to generate Fig. 9 are reported in the 7th column of Table 2.

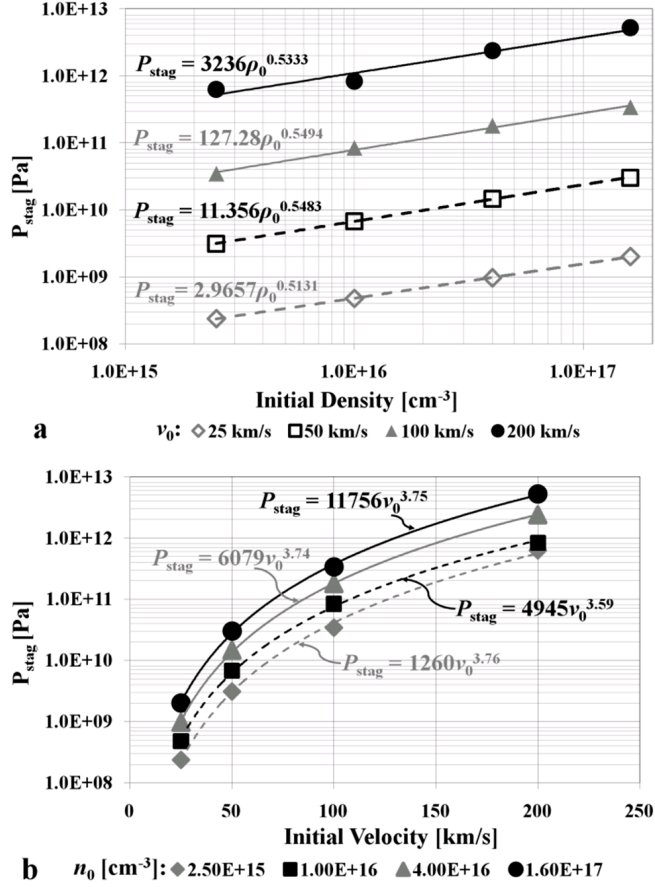

FIG. 9: Stagnation pressure scaling for the 16 simulations defined in Table 2. RAVEN simulations consider Ar liners with $\gamma=5/3$ and $T_0=1.0$ eV. (a) $P_{stag}$ vs. $n_0$ for a given $v_0$. Power law functions are fit to the data, and equations are shown. $P_{stag}$ scales approximately as $n_0^{1/2}$ for a given $v_0$. (b): $P_{stag}$ vs. $v_0$ for a given $n_0$. Power law functions are fit to the data, and equations are shown. $P_{stag}$ scales approximately as $v_0^{15/4}$ for a given $n_0$.

The velocity dependence of the scaling laws derived from the data in Fig 9 is incompatible with steady-state flow solutions. For a very high-$M$ ideal gas with $\gamma=5/3$, the steady-state solution finds that the shock velocity $v_s=v_0/3$ [15]. As discussed in Sec. 3, for the Noh verification tests (where the steady-state flow approximation holds well, and the initial $M$ is infinite) $v_s \sim v_0/3$. If $v_s \propto v_0$, and perfect conversion of liner kinetic energy to stagnated plasma internal energy is assumed, the resultant scaling is $P_{stag} \propto v_0^2$. Liner simulations find (in contrast to steady-state flow solutions) that $v_s$ is quite independent of $v_0$ (and $v_s \ll v_0$, Fig. 10). $v_s$ has been determined by averaging the velocity of shock front from time $t_1=t(p_{max})+\Delta R/5v_0$ to $t_2=t(p_{max})+4\Delta R/5v_0$ (the central 3/5 of $\tau_{stag}$). The low values of $v_s$ are due, in part, to the reduced incoming flow velocity of the fluid adjacent to the shock front. The weak correlation between $v_s$ and $v_0$ is not yet fully understood, and is a topic of ongoing study.

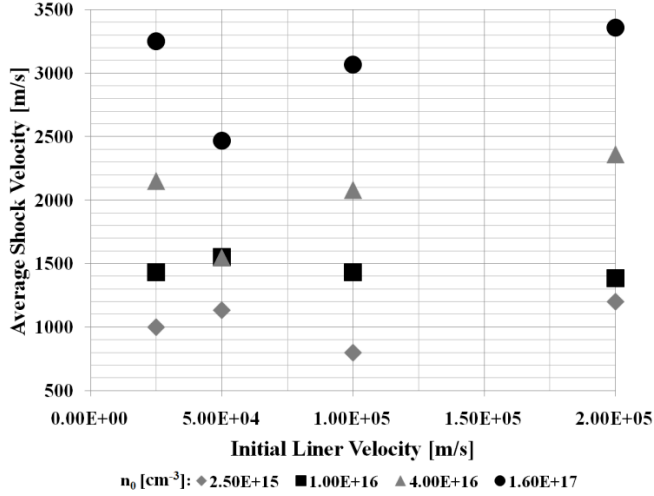

FIG. 10: Average shock velocity ($v_s$, averaged from $t_1=t(p_{max})+\Delta R/5v_0$ to $t_2=t(p_{max})+4\Delta R/5v_0$ for those simulations used to find the $P_{stag}$ scalings shown in Fig. 9. Simulations show that the speed of the outgoing shock is quite independent of the initial incoming flow velocity ($v_0$).

The $P_{stag}\propto v_0^{15/4}$ scaling (Fig. 9) is consistent with the following. Assume, as observed in the simulations, that $v_s$ is both independent of $v_0$ (that is, for a given $n_0$, increasing $v_0$ does not increase $v_s$) and that $v_s\approx$ constant throughout $\tau_{stag}$. Therefore, at $t=t(P_{max})+\tau_{stag}$ the shock front has propagated from the origin to $r_s=v_s\times\tau_{stag}$. Further, assume that all initial liner kinetic energy is deposited in the stagnated plasma. The average energy density (pressure) of the stagnated plasma will then be given by $\varepsilon=KE_0/\{(4/3)\pi(v_s\cdot\tau_{stag})^3\}$. But, $KE_0=\rho_0v_0^2\times(4/3)\pi\{(R_m+\Delta R)^3-R_m^3\}$, and, as shown in Section V.B, $\tau_{stag}=\Delta R/v_0$. Therefore, $\varepsilon=\rho_0v_0^5\{(R_m+\Delta R)^3-R_m^3\}/\{v_s\cdot\Delta R\}^3\propto v_0^5$. This derivation, which assumes no losses (e.g., via radiation), no entropy generation (e.g. by the propagating shock) and perfect conversion of $KE_0$ to stagnated plasma internal energy, provides an upper bound on the velocity scaling. Thus, the observed weaker $v_0^{(15/4)}$ scaling appears reasonable, considering that loss mechanisms are included in the simulations.

## C: Liner thickness effects

The scaling relations described in Sect. VI.B are quite robust over a wide range of $v_0$, $\rho_0$, and $KE_0$ (Table 2), but the simulations used to determine them examine only liners with $R_m$=0.241 m and $\Delta R$=0.255 m. While the $\rho(r)=\rho_0(r_0/r)^2$ density scaling (Sect. IV) can be used to apply the scaling relations to liners with different merging radii, differences in initial liner thickness result in deviations from the scaling law. Whereas increasing the liner thickness does not affect $P_{max}$, $P_{stag}$ is reduced since the pressure of the stagnated plasma decays with time (due, in part, to energy transport), during the plasma confinement time, $\tau_{stag}$ (see Figs. 6, 8). Therefore, even though (e.g.) simulations "PLX 1" and "PLX 2" have precisely the same $\rho_0$, $v_0$, and $R_m$, and nearly identical *early* $P(t, r\sim0)$, due to the increase in $\tau_{stag}$ and the reduced pressure in the latter stages of the stagnation time, the thicker liner (PLX 2) has lower $P_{stag}$; thus, the reported scaling, which have only $\rho_0$ or $v_0$ as input parameters, cannot predict $P_{stag}$ for liners with different $\Delta R$. Practically, if the liner thickness is well defined, similar scaling relationships could be found.

The initial liner thickness will be a critical parameter in a plasma-liner-driven MIF system, since it will affect the fuel pressure, lifetime, and fusion yield. Results have shown that increasing the liner thickness enhances $\tau_{stag}$, but reduces $P_{stag}$. A fusion-relevant metric for liner performance is the Lawson-like parameter $P_{stag} \times \tau_{stag}$ which has been calculated in Tables 1 and 2. As expected, with other parameters held constant, thicker liners achieve higher $P_{stag} \times \tau_{stag}$, but at the cost of increased investment in $KE_0$. An even more meaningful measure of liner performance is the quantity $P_{stag} \times \tau_{stag}/KE_0$ (also included in the tables); thin liners outperform thick liners, again, due to the reduction in $P_{stag}$ associated with longer stagnation times. To provide a quantitative example, simulation 15 in Table 2 was repeated for varying $\Delta R$ of 0.255, 0.216, 0.167, and 0.104 m, with $KE_0$ of 24, 18, 12, and 6 MJ, respectively. Simulation results are summarized in Table 3, and suggest (along with results in Table 1 and 2) that thinner, higher-velocity liners optimize $P_{stag} \times \tau_{stag}$ for a given $KE_0$.

Table 3: Changes in performance with varying initial liner thickness. RAVEN simulations consider $T_0$=1.0 eV argon plasma and use an ideal gas EOS with $\gamma$=5/3. Similar to simulation 15 in Table 2, $R_m$=0.241 m, $n_0$=1.6×10$^{17}$, and $v_0$=100 km/s. Simulations examine varying $\Delta R$ of 0.255, 0.216, 0.167, and 0.104 m, with $KE_0$ of 24, 18, 12, and 6 MJ, respectively. Based on the parameter $P_{stag}\times\tau_{stag}/KE_0$, thin liners outperform thick liners.

|  | Case 1 | Case 2 | Case 3 | Case 4 |
|---|---|---|---|---|
| $\Delta R$ [m] | 0.255 | 0.216 | 0.167 | 0.104 |
| $KE_0$ [J] | 2.4×10$^7$ | 1.8×10$^7$ | 1.2×10$^7$ | 6.0×10$^6$ |
| $P_{max}$ [Pa] | 4.25E+12 | 4.25E+12 | 4.22E+12 | 4.20E+12 |
| $P_{stag}$ [Pa] | 3.21E+11 | 3.41E+11 | 3.74E+11 | 5.07E+11 |
| $\tau_{stag}$ [s] | 2.34E-06 | 1.94E-06 | 1.4E-06 | 6.9E-07 |
| $P_{stag}\times\tau_{stag}$ | 750600 | 661012 | 524295 | 350028 |
| $P_{stag}\times\tau_{stag}/KE_0$ | 0.031 | 0.037 | 0.044 | 0.058 |

**D: Stagnation pressure scaling with Initial Mach number**

To investigate $P_{stag}$ scaling with *initial* liner Mach number ($M_0$), the series of 16 simulations in Table 2 were again run with varying atomic species/mass, $\gamma$, and $T_0$. Three additional series were considered with varying liner plasmas: (1) Ar plasma with $\gamma$=5/3 and $T_0$=10.0 eV; (2) Ar plasma with $\gamma$=1.1 and $T_0$=1.0 eV; (3) Xe plasma with $\gamma$=5/3 and $T_0$=1.0. For Xe simulations, $\rho_0$ and $KE_0$ equal those values quoted in Table 2; thus $n_0$ is reduced by the atomic-mass ratio of Ar to Xe (39.948/131.29). By the same method described in Section V.C, $P_{stag}$ has been calculated for each of the additional 48 runs. A normalized pressure ($P_n$) is found by dividing $P_{stag}$ by $\rho_0 v_0^2/2$. $P_n$ is then plotted (Fig. 11) versus $M_0$=$v_0/c_s$=$v_0(m/\gamma k T_0)^{1/2}$ (where $m$ is the atomic mass), to incorporate the three new variables under consideration ($\gamma$, $T_0$, and $m$,). In the figure, data are grouped according to the simulation atomic species, $\gamma$, and $T_0$ by the dashed rectangles (Ar, $\gamma=5/3$, $T_0=1.0$ eV), solid ovals (Xe, $\gamma=5/3$, $T_0=1.0$ eV), solid rectangles (Ar, $\gamma=5/3$, $T_0=10$ eV), and dashed ovals (Ar, $\gamma=1.1$, $T_0=1.0$ eV). Finally, the surrounding rectangles and ovals also indicate the initial velocity since four of each marker type (e.g., solid rectangles) exist on the plot, with the leftmost corresponding to $v_0=25$ km/s, the second from the left to $v_0=50$ km/s, the second from the right to $v_0=100$ km/s, and the rightmost to $v_0=200$ km/s. These velocities are indicated on the plot only for the Ar, $\gamma=5/3$, $T_0=10$ eV data (solid rectangles).

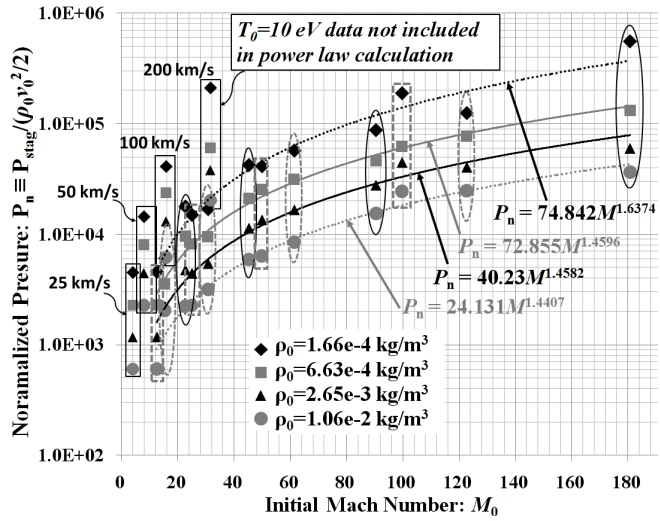

FIG. 11: RAVEN calculated normalized pressure {$P_n=P_{stag}/(\rho_0 v_0^2/2)$}versus initial Mach number ($M_0$) for 4 different sets of the 16 simulations defined in Table 2. Data are grouped according to the simulation atomic species, $\gamma$, and $T_0$ by the dashed rectangles (Ar, $\gamma=5/3$, $T_0=1.0$ eV), solid ovals (Xe, $\gamma=5/3$, $T_0=1.0$ eV), solid rectangles (Ar, $\gamma=5/3$, $T_0=10$ eV), and dashed ovals (Ar, $\gamma=1.1$, $T_0=1.0$ eV). Separate scaling laws have been fit to the $T_0=1.0$ eV data for each initial density. Power law scaling functions are best fit to the data, and given by $P_n \propto M_0^{1.64}$, $P_n \propto M_0^{1.46}$, $P_n \propto M_0^{1.46}$, and $P_n \propto M_0^{1.44}$, for $\rho_0=1.66\times10^{-4}$, $6.63\times10^{-4}$, $2.65\times10^{-3}$, and $1.06\times10^{-2}$ kg/m$^3$, respectively.

Data in Fig. 11 show strong correlation between $M_0$ and $P_n$, yet a single scaling law for all data is not adequate. First, for a given $v_0$, an increase in $\rho_0$ increases $P_n$ without altering $M_0$. Second, there is a clear disparity between those simulations with $T_0=10$ eV (solid rectangles) and the remaining 1.0 eV simulations. The higher (10 eV) temperature reduces $M_0$, however, during convergence, the higher-temperature high-Z liners radiate, and the plasma temperature quickly falls. By the time of void closure, the temperature difference between similar simulations with $T_0$=10 eV and $T_0$=1 eV is minimal, and therefore the achieved stagnation pressures are comparable. Also, while such effects are strongest at low $v_0$, decreasing the adiabatic constant increases the stagnation pressure for similar $n_0$ and $v_0$. Separate scaling laws have been fit to the $T_0$=1.0 eV data only, for each initial density (Fig. 11). Normalized stagnation pressures $P_n$={$P_{stag}/(\rho_0 v_0^2/2)$} are best fit to $P_n \propto M_0^{1.64}$, $P_n \propto M_0^{1.46}$, $P_n \propto M_0^{1.46}$, and $P_n \propto M_0^{1.44}$, for $\rho_0$=1.66×10$^{-4}$, 6.63×10$^{-4}$, 2.65×10$^{-3}$, and 1.06×10$^{-2}$ kg/m$^3$, respectively. These scaling, along with those presented in Section V.C, suggest that the initial flow velocity is the dominant factor in achieving high stagnation pressures.

It is interesting to note that stagnation pressure scaling with Mach number was also examined previously both computationally [17] and analytically [18] for inertial confinement fusion (ICF) relevant implosions. Despite the drastic differences in model sophistication in those studies (2D radiation-hydrodynamics simulations including transport and EOS/opacity versus self-similar analytic solutions of 1D ideal hydrodynamics), they agreed that the *peak pressure* after void closure ($P_{peak}$) of an imploding spherical shell scales as $P_{peak}/P_d \propto M_c^{8/(\gamma+1)}=M_c^3$ for $\gamma$=5/3, where $P_d$ is the maximum drive pressure applied to the shell (i.e., by a high-power laser), and $M_c$ is the spatially-uniform Mach number *at the time of void closure*. While clear differences exist between ICF-shell implosions and plasma liner implosions (most notably, the high-impulse laser driver launches a shock which creates a positive radial density gradient at void closure, whereas the initial conditions of uniformly distributed kinetic energy in a plasma liner lead to a shock-free convergent flow, and a negative radial density gradient at void closure), plasma-liner Mach number scaling is nonetheless observed (Fig. 11).

### E: Remarks on equation-of-state modeling

All of the RAVEN results reported in this paper are based on an ideal gas, constant-$\gamma$ EOS, which neglects some potentially important physical effects needed for the accurate modeling of imploding plasma liners. Perhaps most important among these are: (1) contribution of the (time-dependent) ionization state ($Z_{\text{eff}}$) to total plasma pressure, and (2) density and temperature dependent EOS that accounts for plasma internal degrees of freedom associated with atomic physics effects such as ionization, line transitions/radiation, and recombination. The latter has the potential for keeping the liner entropy low during convergence, thereby enabling it to reach higher $P_{\text{stag}}$. Both effects are expected to alter the accuracy of the hydrodynamic modeling, but by how much and to what extent does an ideal gas EOS over- or underestimate $P_{\text{stag}}$ remains uncertain. It has been suggested that setting $\gamma < 5/3$ could capture some of the effects of a more realistic EOS model [19]. Using HELIOS, indeed it was observed that setting $\gamma = 1.2$ and including ionization effects (i.e., enabling a time-dependent $Z_{\text{eff}}$) provided the best match to simulations performed using the PROPACEOS non-LTE EOS table (Fig. 12). However, a more detailed and systematic study is needed to fully assess the importance of EOS models for imploding plasma liners. What is needed is the ability to independently choose a given plasma liner initial condition (including its EOS and $Z_{\text{eff}}$) and then evolve the liner using different EOS models. This is deferred to future work.

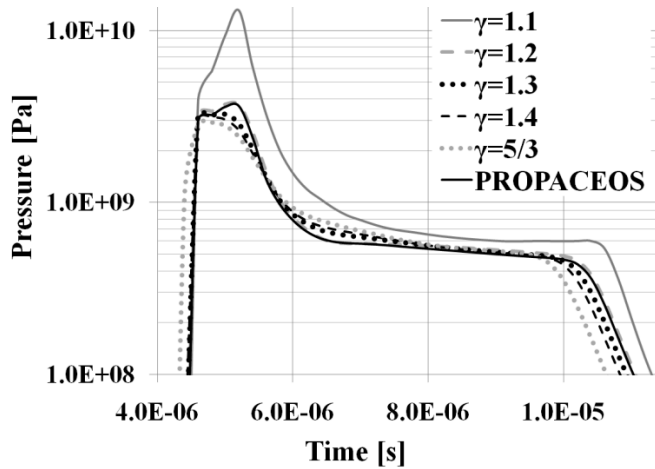

FIG. 12: HELIOS calculated $P(t)$ curves (case 6 of Table 2) from 6 separate simulations. The first five simulations use an ideal gas EOS with ionization (i.e., a time-dependent $Z_{\text{eff}}$). The adiabatic index, $\gamma$, is initially set to 1.1, 1.2, 1.3, 1.4, or 5/3. The sixth simulation uses the PROPACEOS non-LTE EOS. Results from the ideal gas EOS simulations with $\gamma = 1.2$ match well with PROPACEOS results.

## VII: CONCLUDING REMARKS

One-dimensional radiation-hydrodynamics simulations have been performed to examine the scaling of stagnation pressure as a function of the initial conditions of imploding, spherically symmetric plasma liners. Simulations (which ignore multi-dimensional effects but include radiation transport and thermal conduction) using liner parameters which may be achieved by PLX suggest that plasmas with stagnation pressures near 1 Mbar can be sustained for 1 µs, or that pressures near 10 kBar can be sustained for 10 µs. It is shown that radiation and thermal conduction must be included to avoid the formation of an unphysically-high-temperature "plasma bubble" which artificially limits the convergence and peak pressure achieved by the imploding plasma liner. By examining a variety of liners with parameters outside of those accessible to PLX, scaling laws for higher-energy liners have been obtained, which can be used to evaluate the plasma-liner concept for HED physics and MIF applications. Over the parameter space investigated, $P_{\text{stag}}$ scales approximately as $n_0^{1/2}$ for a given $v_0$ and as $v_0^{15/4}$ for a given $n_0$. Such strong velocity scaling is realistic only if the shock-front speed is both considerably less than and relatively independent of the incoming flow speed. Such shock characteristics are observed in simulations. It is interesting to note that based on the $P_{\text{stag}} \sim v_0^{15/4}$ scaling derived from these 1D simulations, MIF relevant pressures near 50 Mbar may be sustained for nearly 0.6 µs with an argon plasma liner with $v_0 \sim 150$ km/s and $KE_0 \sim 50$ MJ (these requirements will likely increase when multi-dimensional effects are included). Normalized stagnation pressure ($P_n$) scaling with initial liner Mach number ($M_0$) is approximately $P_n \propto M_0^{1.5}$ over varying atomic species, adiabatic index, and initial plasma density. Computational results suggest that experimental data will provide a unique and rich dataset for validating transport and EOS models.

For assistance with RAVEN simulations, we thank W. Atchison, A. Kaul, and C. Rousculp. For assistance with HELIOS simulations, we thank J. MacFarlane and Prism Computational Sciences, Inc. Finally, we thank B. Bauer, G. Kagan, M. Stanic, X. Tang, Y. C. F. Thio, and F. D. Witherspoon for many useful conversations. This work was supported by the Office of Fusion Energy Sciences of the U.S. Department of Energy under contract No. DE-AC52-06NA25396 and a National Undergraduate Fellowship in Plasma Physics and Fusion Energy Sciences (JSD).

[1] I. R. Lindemuth and R. C. Kirkpatrick, Nucl. Fusion **23**, 263 (1983).

[2] R. Kirkpatrick, I. R. Lindemuth, and M. S. Ward, Fusion Tech. **27**, 201 (1995)

[3] I.R. Lindemuth and R.E. Siemon, Am. J. Phys. **77,** 409 (2009).

[4] Y. C. F. Thio, E. Panarella, R. C. Kirkpatrick, C. E. Knapp, F. Wysocki, P. Parks, and G. Schmidt, "Magnetized target fusion in a spheroidal geometry with standoff drivers," in *Current Trends in International Fusion Research II*, edited by E. Panarella (National Research Council Canada, Ottawa, Canada, 1999).

[5] S. C. Hsu, T. J. Awe, D. S. Hanna, J. S. Davis, F. D. Witherspoon, J. T. Cassibry, M. A. Gilmore, D. Q. Hwang, and the PLX Team, Bull. Amer. Phys. Soc. **55**, 357 (2010).

[6] F. D. Witherspoon, R. Bomgardner, A. Case, S. Messer, S. Brockington, L. Wu, R. Elton, S. Hsu, J. Cassibry, M. Gilmore, and the PLX Team, Bull. Amer. Phys. Soc. **55**, 358 (2010).

[7] D. D. Ryutov, "Using plasma jets to simulate galactic outflows," Plasma Jet Workshop, Los Alamos, January 24-25 (2008).

[8] J. T. Cassibry, M. D. Stanic, T. J. Awe, D. S. Hanna, J. S. Davis, S. C. Hsu, F. D. Witherspoon, Bull. Amer. Phys. Soc. **55**, 359 (2010).

[9] C. Thoma, D. Welch, R. Clark, J. MacFarlane, I. Golovkin, and F. D. Witherspoon, Bull. Amer. Phys. Soc. **55**, 360 (2010); J. R. Thompson, N. I. Bogatu, S. A. Galkin, J. S. Kim, D. R. Welch, C. Thoma, J. J. MacFarlane, F. D. Witherspoon, J. T. Cassibry, T. J. Awe, S. C. Hsu, *ibid* **55**, 360 (2010).

[10] R.J. Kanzleiter, W.L. Atchison, R.L. Bowers, R.L. Fortson, J.A. Guzik, R.T. Olson, J.L. Stokes, and P.J. Turchi, IEEE Trans. Plasma Sci., **30**, 1755 (2002).

[11] See http://www.prism-cs.com/Software/Helios/Helios.htm for further information about HELIOS.

[12] See http://www.prism-cs.com/Software/PROPACEOS/PROPACEOS.htm for further information about PROPACEOS.

[13] The table was generated by PRISM Computational Sciences to include densities as low as $10^{10}$ cm$^{-3}$ and temperatures as low as 0.01 eV for our use.

[14] W.F. Noh, J. Computational Phys. **72**, 78 (1987).

[15] P. B. Parks, Phys. Plasmas **15**, 062506 (2008).

[16] R. Samulyak, P. Parks, and L. Wu, Phys. Plasmas **17**, 092702 (2010).

[17] M. C. Herrmann, M. Tabak, and J. D. Lindl, Phys. Plasmas **8**, 2296 (2001).

[18] A. Kemp, J. Meyer-ter-Vehn, and S. Atzeni, Phys. Rev. Lett. **86**, 3336–3339 (2001).

[19] M. Murakami and K. Nishihara, Phys. Plasmas **7**, 2978, 2000.